\documentclass[11pt,a4paper]{article}
\usepackage{authblk}
\usepackage{amsmath}
\usepackage{amssymb}
\usepackage{amsthm}
\usepackage{amsfonts}
\usepackage{enumerate}
\usepackage{pstricks}
\usepackage{pst-plot}
\usepackage{pst-poly}
\usepackage{graphicx}
\usepackage{mathtools} 
\usepackage{url}
\usepackage{booktabs}
\usepackage{multirow}
\usepackage{verbatim}
\usepackage{hyperref}
\usepackage{breakurl,url}
\usepackage[margin=1.15in]{geometry}
\usepackage{mathrsfs}%
\usepackage[title]{appendix}%
\usepackage{xcolor}%
\usepackage{textcomp}%
\usepackage{manyfoot}%
\usepackage{algorithm}%
\usepackage{algorithmicx}%
\usepackage{algpseudocode}%
\usepackage{listings}%
\usepackage{todonotes}

\theoremstyle{thmstyleone}%
\theoremstyle{thmstyletwo}%

\theoremstyle{thmstylethree}%

\newcommand{\R}[0]{{\mathbb{R}}}
\DeclareMathOperator{\Aut}{Aut}
\DeclareMathOperator{\tr}{tr}
\DeclareMathOperator{\diag}{diag}
\DeclareMathOperator{\LR}{LR}   
\DeclareMathOperator{\HD}{d^H}   
\DeclareMathOperator{\IP}{P_{\text{init}}}   
\DeclareMathOperator{\FinP}{P_{\text{fin}}}   
\DeclareMathOperator{\OPS}{\mathcal{P}_{\text{opt}}}   

\makeatletter
	\newcommand\myparagraph{%
    \@startsection{paragraph}{4}{0mm}%
        {-\baselineskip}%
		{-0.7\baselineskip}%
        {\normalfont\normalsize\bfseries}}
\makeatother

\begin{document}

\title{Computing approximate symmetries of complex networks}

\author[1,2]{Anna Pidnebesna}

\author[1,3]{David Hartman}

\author[1,3]{Aneta Pokorn\'{a}}

\author[3]{Mat\v{e}j Straka}

\author[1,2]{Jaroslav Hlinka}

\affil[1]{Institute of Computer Science of the Czech Academy of Sciences, Czech Academy of Sciences, Pod Vod\'{a}renskou v\v{e}\v{z}\'{i} 271/2, Prague 8, 18207, Czech Republic}

\affil[2]{National Institute of Mental Health, Topolov\'{a} 748, Prague, 25067, Czech Republic}

\affil[3]{Computer Science Institute of Charles University, Faculty of Mathematics and Physics, Charles University, Malostransk\'{e} n\'{a}m. 25, Prague 1, 11800, Czech Republic}

\date{}
\maketitle

\begin{abstract}
The symmetry of complex networks is a global property that has recently gained attention since MacArthur et al. 2008 showed that many real-world networks contain a considerable number of symmetries. These authors work with a very strict symmetry definition based on the network's automorphism. The potential problem with this approach is that even a slight change in the graph's structure can remove or create some symmetry. Recently, Liu 2020 proposed to use an approximate automorphism instead of strict automorphism. This method can discover symmetries in the network while accepting some minor imperfections in their structure. The proposed numerical method, however, exhibits some performance problems and has some limitations while it assumes the absence of fixed points. In this work, we exploit alternative approaches recently developed for treating the Graph Matching Problem and propose a method, which we will refer to as Quadratic Symmetry Approximator (QSA), to address the aforementioned shortcomings. To test our method, we propose a set of random graph models suitable for assessing a wide family of approximate symmetry algorithms. The performance of our method is also demonstrated on brain networks. 
\end{abstract}


\section{Introduction}

Complex networks represent a modern technique for analyzing dynamic systems, which allows the analysis of systems that consist of many interconnected and interacting elements~\cite{barabasi2016network,bullmore2009complex,boccaletti2006complex}. Various characteristics of networks can contribute significantly to the analysis of such systems. These characteristics are studied either at the local level, such as identifying hubs~\cite{van2013network,rubinov2010complex}, or at the global level, such as the small-world character~\cite{watts1998collective}. Any characteristic may have problems with application in complex networks due to its sensitivity to the specific network structure; see the problems in the definition of the small-world coefficient~\cite{hlinka2012small,hlinka2017small}.

Recently, network symmetry has proven to be an important property studied within complex networks~\cite{macarthur2008symmetry,sanchez2020exploiting,della2020symmetries}. Most symmetry characteristics are based on the automorphism group of the corresponding graph. This can be called a global characteristic of the graph, i.e. a characteristic that evaluates the entire graph. As noted by S\'{a}nchez-Garc\'{i}a~\cite{sanchez2020exploiting}, network symmetries are further inherited by any characteristic of the network. For that reason, symmetries are also studied locally for selected characteristics, such as the betweenness centrality~\cite{gago2013betweenness,hartman2024ontheconnectivity,ghanbari2023structure}. Both approaches to symmetries from a local or global perspective face problems with their rigorous definition. The challenge here is that the definition of a network can be burdened with uncertainty. The above-given global symmetry characteristic, based on automorphisms, is quite sensitive to small changes in graph structures. Alternatively, by the nature of the system, the strict symmetry condition may be too strong in a complex network based on a real system.

Recently, it has been preliminarily suggested to use an alternative notion of approximate network symmetry~\cite{liu2020approximate}. This symmetry also considers the automorphism group but allows a small number of edges to break the isomorphism condition. The authors of the above-mentioned work suggested using simulated annealing to find the most suitable permutation that minimizes the number of symmetry-violating pairs of vertices. However, this approach has several drawbacks, such as the balance between the symmetry of the obtained solutions and the corresponding computational complexity or the inability of the suggested method to allow fixed points in the respective permutation. 

Searching for approximate automorphisms is closely related to the graph matching problem (GMP), which is known to be equivalent to a quadratic assignment problem (QAP). Recently, Vogelstein et al.~\cite{vogelstein2015fast} suggested a relaxation method to solve QAP. For such a relaxation, it is possible to construct an iterative method that makes use of approaches from standard nonlinear solvers. 

In this work, we suggest a modification of the previously mentioned relaxed QAP to a relaxed approximate symmetry problem (rASP) that is able to evaluate the approximate network symmetry. The original relaxed QAP method allows having fixed points in the nodes' permutations. When computing the symmetry of a single graph, this naive method could converge to the identity solution, which is not desirable for our purposes (of finding non-trivial permutations). Thus, we penalize the number of fixed points in the suggested solution, finding approximately symmetrical non-identical permutations. For the numerical evaluation of rASP we use the modified algorithm of Vogelstein et al.~\cite{vogelstein2015fast}, which we call Quadratic Symmetry Approximator (QSA). Using QSA and simulated annealing algorithms, we perform extensive numerical simulations on random networks to compare these approaches. Compared to the simulated annealing approach, the suggested method QSA significantly improves the symmetry measure of the solutions. 

The paper is organized as follows. Section~\ref{sec:network_symmetry} contains the main definitions considering the network symmetry, basic notation of corresponding areas such as graph theory, the definition of approximate network symmetries, and the design of respective algorithms. Section~\ref{sec:data} describes various models used later to test and compare the approaches and describes the real data used in our study of network approximate symmetries in real-world settings. Section~\ref{sec:results} is devoted to the numerical results.

\section{Network symmetry}
\label{sec:network_symmetry}

This section reviews the basic notation and knowledge of network symmetry and its approximate version. It also quickly introduces some areas necessary for defining this approach, such as graph isomorphism. We use standard notation for graph theory and linear algebra, as in~\cite{godsil2001algebraic}.

\subsection{Basic notation}

For matrices, we consider the standard notation.  $I_n$ is the identity matrix of size $n \times n$. The operator $tr(A)$ denotes the trace of the matrix $A$, i.e. the sum of the elements on the diagonal, $e = (1, \ldots ,1)^T$ denotes the vector of ones (with convenient dimension), $diag(v)$ is the diagonal matrix with entries given by vector $v$.

The main object of interest is the (undirected) graph $G$, defined as the pair $G = (V, E)$. The set $V = \{v_1, \ldots, v_n\}$ encloses nodes (or vertices) and the set $E$ contains edges. Each edge can be considered as a two-element node set that we call endpoints, e.g. an edge $\{v_2, v_3\}$. We do not consider loops, i.e. edges having both endpoints the same. For simplicity, we denote by $n = |V|$ and $m = |E|$. Given a graph $G$ we can encode its structure via an adjacency matrix $A$ defined as $A_{ij}$ equal to $1$ if $\{v_i, v_j\}\in E(G)$ and $0$ otherwise.

Consider two graphs $G$ and $H$. A bijective mapping $f:V(G) \rightarrow V(H)$ is called an isomorphism if $\{v,v'\} \in E(G) \Leftrightarrow \{f(v),f(v')\} \in E(G)$ holds for any $\{v,v'\} \subseteq V(G)$. We can assume that the sizes of the graphs are the same and, in addition, that both graphs have a given ordering of vertices, e.g., $V(G) = \{v_1, \ldots, v_n\}$ and $V(H) = \{u_1, \ldots, u_n\}$. Consequently, isomorphism can be viewed as a reshuffling of the vertices of the first graph so that the graphs are identical (edges as well as non-edges are kept). A permutation can represent this reshuffling. We often refer to a permutation as $\pi$ and to the set of all permutations of an $n$-element set as $S_n$. We can encode permutation $\pi\in S_n$ via the permutation matrix $P\in \{0,1\}^{n\times n}$ for which $P_{ij} = 1$ if and only if $\pi(i) = j$. The set of all permutation matrices will be denoted as $\mathcal{P}_n$. The problem of finding a graph isomorphism can be thus naturally translated into the language of matrices as follows:
\begin{align*}
&\text{Graph isomorphism problem (GI)} \\
&\quad \text{{\em Input:} Two graphs $G, H$ having adjacency matrices } A, B	\\
&\quad \text{{\em Task:} Find permutation matrix $P$ s.t } A = PBP^T
\end{align*}

A special case of isomorphism is when we consider a mapping of a graph onto itself. More formally, an automorphism is an isomorphism $f: V(G) \rightarrow V(G)$. Again, automorphisms can be viewed as permutations. The $\pi(i) = j$ denotes the situation where a node $v_i$ is mapped to a node $v_j$. A special situation occurs when the permutation $\pi$ maps a particular node $v_i$ to itself, i.e. $\pi(i) = i$. We refer to this node $v_i$ as a fixed point. All automorphisms of a given graph are grouped in the so-called automorphism group $\Aut(G)$, which can be thought of as a set of all automorphisms on $G$ for our purposes, i.e. ignoring the group additional structure~\cite{godsil2001algebraic}. Note that the maximal possible number of automorphisms equals the maximal number of permutations, i.e., $n!$. This many automorphisms can be achieved for complete graphs in which all edges are present or empty graphs without any edge. Each graph also has a trivial automorphism (or identity) that maps each node to itself. Notably, there are graphs, called asymmetric, in which one cannot find any non-trivial automorphism. In fact, those graphs are surprisingly frequent~\cite{erdos1963asymmetric}.

\subsection{Approximate network symmetries}

One way around the disadvantages of strict symmetries is to allow disagreement for several pairs of nodes. That is, we declare a given mapping to be an approximate symmetry if it satisfies the definition of automorphism except for a few pairs of vertices. Following~\cite{liu2020approximate}, the number of disagreeing pairs of nodes for a given permutation $P$ can be computed as 
\begin{equation*}
\epsilon(A,P) = \frac{1}{4} \|A - PAP^T\|_F,
\end{equation*}
Note that any permutation automatically keeps the total number of edges. Then, the constant $1/4$ is chosen so that the resulting norm gives the number of distinct pairs. Such a choice has two reasons: 1) once an edge is missing somewhere in the image, it must be present elsewhere, and 2) the adjacency matrix represents each pair twice. The following minimization problem then describes the task of finding the best approximate automorphism.
\begin{equation*}
E(A) = \min_{P\in \mathcal{P}_n} \epsilon(A,P) = \min_{P\in \mathcal{P}_n}\frac{1}{4} \|A - PAP^T\|_F
\end{equation*}

Obviously, the value $E(A) = 0$ corresponds to a strict symmetry. On the other hand, the maximum value is ${n \choose 2}$, which depends on the size of the network, so a normalized version for this coefficient was introduced~\cite{liu2020approximate}.
\begin{equation}
S(A) = \frac{E(A)}{\frac{1}{2}{n \choose 2}} = \frac{\|A - PAP^T\|_F}{n(n-1)}
\end{equation}

We will refer to this characteristic as the (normalized) approximate symmetry coefficient. There are two problems with this definition. First, the identity is always a solution. Second, the number of permutations is large, and attempts to find the minimum can be difficult. Liu proposed a solution to the first problem and made a step towards solving the second problem by restricting to permutations without fixed points. This step, of course, disqualifies the identity. Unfortunately, the second problem remains since the number of permutations without a fixed point is still approximately $n!/e$. 

As mentioned above, computing the minimum of $\epsilon(A,P)$ is challenging due to the size of the space $\mathcal{P}_n$. The original method of Liu~\cite{liu2020approximate} chosen for this purpose was simulated annealing with the logarithmic cooling scheme $T(t) = \frac{c}{\log(t+d)}$.

Note also that not allowing any fixed points in the resulting permutation can be a substantial restriction of the problem space evaluated by the algorithm, which could lead to unsatisfying results. Thus, our study uses the modified version of the algorithm, which allows up to $K$ fixed points in a solution, where $K \in \mathrm{N_0}$ is a parameter. We call this version of an algorithm {\em Annealing with Fixed Points} (AFP).

\subsection{Design of method based on quadratic assignment}

The problem of finding automorphisms is closely related to the {\em Graph isomorphism problem} (GI), the problem of finding the isomorphism of two graphs. Since we aim to find approximate automorphisms, we will be concerned with the problem of finding inexact isomorphisms. This problem can be expressed using the {\em Graph Matching Problem}(GMP)~\cite{conte2004thirty,livi2013graph}. 
\begin{align*}
&\text{Graph matching problem (GMP)} \\
&\quad \text{{\em Input:} Two graphs $G, H$ having adjacency matrices } A, B	\\
&\quad \text{{\em Task:} } \min_{P\in \mathcal{P}_n} \|A - PBP^T\|_F	
\end{align*}

There are several methods for solving this problem. We stay with the isomorphism perspective and consider one way to solve the original problem. The problem of finding isomorphism can be expressed using an existing problem called the {\em Quadratic assignment problem} (QAP)~\cite{koopmans1957assignment}. This problem considers $n$ facilities and $n$ locations, for which we have distances between all locations, represented by a distance matrix $D \in \R^{n\times n}$, and requested transportation between facilities, represented by a flow matrix $F \in \R^{n\times n}$. The challenge entails allocating the facilities to distinct locations with the objective of minimizing the total product of distances and corresponding flows, that is, the transportation costs.
The problem can then be represented in a matrix form as the following minimization problem~\cite{koopmans1957assignment}.
\begin{align*}
&\text{Quadratic assignment problem (QAP)} \\
&\quad \text{{\em Input:} Matrix $F \in \R^{n\times n}$ of flows and matrix $D \in \R^{n\times n}$ of distances}\\
&\quad \text{{\em Task:} } \min_{P\in \mathcal{P}_n} tr(FPD^TP^T)
\end{align*}

If we realize that vertex mapping by isomorphism is close to assignment in the context of the Quadratic assignment problem, we can easily see that the two problems are deeply related. If we follow the steps from~\cite{vogelstein2015fast} we get the Graph matching problem in the following form.
\begin{align*}
&\text{Graph matching problem (GMP) - redefined} \\
&\quad \text{{\em Input:} Two graphs $G, H$ having adjacency matrices } A, B	\\
&\quad \text{{\em Task:} } \min_{P\in \mathcal{P}_n} -\tr(APB^TP^T)
\end{align*}

The problem of seeking the best approximate graph symmetry, i.e. the best approximate (non-trivial) automorphism, can be then formulated as: 

\begin{align}\label{eq:GASP}
&\text{(General) Approximate Symmetry Problem (ASP)} \nonumber\\
&\quad \text{{\em Input:} A graph $G$ having adjacency matrix $A$ } 	\\
&\quad \text{{\em Task:} } \min_{P\in \mathcal{P}_n \setminus \{I_n\}} - \tr(APA^TP^T)\nonumber
\end{align}

Note that while the GMP considers all possible permutations, here we want to use it for finding optimal non-trivial automorphism, and thus, we have to avoid the permutation $P = I_n$, which denotes identity. 

A practical alternative to removing the identity explicitly from the search space is to use a penalization. We introduce the $n$ coefficients $c \in \R^{n}$ and finally introduce the (penalized) approximate symmetry problem as follows.

\begin{align}\label{eq:ASP}
&\text{Approximate symmetry problem (ASP)} \nonumber\\
&\quad \text{{\em Input:} A graph $G$ having adjacency matrix $A$  and a (vector) constant $c$ }	\\
&\quad \text{{\em Task:} } \min_{P\in \mathcal{P}_n} - \tr(APA^TP^T - \diag(c)P)\nonumber
\end{align}

By setting the $c$ parameters appropriately, it is possible to penalize or enforce the use of fixed points.

\subsection{Numerical solution design}

The approximate symmetry problem defined by~\eqref{eq:ASP}, as well as other related problems such as QAP or GMP, have a major obstacle in their solution, which is the combinatorial character of the space $\mathcal{P}_n$. To solve this problem, we can use a relaxation method applied to QAP based on finding the convex envelope of $\mathcal{P}_n$~\cite{vogelstein2015fast}. The idea is to use the doubly stochastic matrices instead of the permutation matrices. A doubly stochastic matrix has all its entries of non-negative, and the sum of each column, as well as the sum of each row, is one. The corresponding space $\mathcal{D}_n$ is the space of all non-negative matrices $P\in \R^{n\times n}$ such that $P^Te = Pe = e$. Analogously to the relaxation applied to the QAP~\cite{vogelstein2015fast}, we perform the same procedure for ASP:

\begin{align*}
&\text{Relaxed approximate symmetry problem (rASP)} \\
&\quad \text{{\em Input:} A graph $G$ having adjacency matrix } A	\\
&\quad \text{{\em Task:} } \min_{P\in \mathcal{D}_n} - \tr(APA^TP^T - \diag(c)P)
\end{align*}

We can also adopt the idea of the solution from~\cite{vogelstein2015fast}. The basic steps are: 1) initialize the solver using any permutation, 2) find a local solution to the problem, and 3) project the doubly stochastic matrix to the nearest permutation matrix. We call this approach the Quadratic Symmetry Approximator (QSA).

\section{Data and models}
\label{sec:data}

This section describes the data and models used to test the above-proposed procedure. 

\subsection{Models}\label{s:models}
As discussed in the sections above, network symmetries can be observed in many systems modeled by complex networks. The same can be expected for approximate symmetries. For this reason, it is useful to test the performance of the algorithms first for different random network models expressing different limiting properties before applying the approach to real-world data. 

\myparagraph{Grid}
The first graph model is not an essentially random graph. It is a class of graphs parameterized by increasing size, called grid graph $R_{n_1, n_2, \ldots, n_d}$. For two dimensions, i.e. $d = 2$, this graph is characterized by the fact that it can be drawn in the plane as a rectangular grid of dimensions $n_1$ and $n_2$ (thus having $n_1n_2$ vertices). This graph can be in full generality defined using graph products~\cite{hahn2013graph}. For two graphs $G$ and $H$, we call a graph $G \times H$ a Cartesian product of $G$ and $H$ if 1) its vertex set is $V(G \times H) = V(G) \times V(H)$, and 2) the vertices $(u,u'), (v,v') \in V(G \times H)$ are adjacent if $u = v$ and $\{u',v'\} \in E(H)$ or $u' = v'$ and $\{u,v\} \in E(G)$. The grid graph $R_{n_1, n_2, \ldots, n_d}$ is then defined as the following product of multiple path graphs $P_{n_1} \times P_{n_2} \times \ldots \times P_{n_d}$. The value of $d$ can be seen as the dimensionality of the grid. For example, you can define a 2D grid ($d=2$) that can be drawn on a plane or a 3D grid ($d=3$) that can be drawn in 3D space.

\myparagraph{Erd\H{o}s-Renyi model}
The first here considered truly random model is the standard {\em Erd\H{o}s-Renyi} (ER) graph model~\cite{erdds1959random}. This model, denoted as $G_{n,p}$, is generated on the node set $V = \{v_1, v_2, \ldots, v_n\}$. An edge is added with probability $p$ for each pair of nodes. This model is not a very accurate representation of complex networks, but it represents a critical limiting behavior object for various properties, such as small-world property.

\myparagraph{Barabási-Albert model} 
A more realistic random model is the Barabási-Albert model~\cite{barabasi1999emergence} (BA). Graphs generated from this model more closely correspond to real complex networks by possessing the scale-free degree distribution, so common in real-world systems~\cite{barabasi1999emergence}. The generative process starts with a small connected graph.  Then, we iteratively add nodes and increase the size of the generated graph (up to a size $n$). Each new node is connected to a limited number of nodes (parameter of BA model) in the existing graph, prioritizing nodes with high degrees (preferential attachment). It is known that the BA model is not perfect in reproducing strict symmetries~\cite{xiao2008emergence}. Nevertheless, it is suitable as an essential representation of complex networks regarding their degree distribution and connectivity of important nodes~\cite{barabasi2016network}.

\myparagraph{Stochastic Block Model}
The random models described above are not appropriate models for representing networks that possess substantial community structure, such as the brain. For this reason, various models have been developed to represent this phenomenon. One of the models is the {\em stochastic block model} (SBM)~\cite{holland1983stochastic}, generalizing the Erd\H{o}s-Renyi models. Suppose that in the target networks generated by this model, we expect communities $C_1, C_2, \ldots, C_r$ such that the size of the community $C_i$ is denoted by $n_i$ and their aggregate size is $n = n_1 + n_2 + \ldots + n_r$. The basic idea is to generate each community using the ER model (with some probability parameter $p_i$, potentially specific for each community) and further generate their connections randomly using some lower probability (to model the relative sparseness of the connectivity between the modules). For such a model, we need predefined sizes of components $\{n_1, n_2, \ldots, n_r\}$ and a square symmetric matrix $P \in \mathbb{R}^{r\times r}$ whose elements are the corresponding probabilities. The diagonal elements of $P_{ii}$ are the probabilities of the ER model corresponding to the community $C_i$ (generated as standard ER model), and the elements of $P_{ij} = P_{ji}$ are the probabilities of edges connecting two communities $C_i$ and $C_j$ (generated as a random bipartite graph). Note that a typical structure of such a matrix will have large elements on the diagonal and small ones off the diagonal.

\myparagraph{Lateral Random Model}
Some real systems have intrinsic symmetry built right into their definition, such as the lateral symmetry of the brain~\cite{Jajcay2022}. Motivated by brain structure, we propose the following model for testing approximate symmetries. The {\em lateral random model} (LRM) $L_{n,p,q}$ can be generated by the following procedure. Generate $G$ as random Erd\H{o}s-Renyi graph $G_{\frac{n}{2},p}$ and consider its copy $G'$. Assume that the graphs have enumerations of their vertices defined as $V(G) = \{v_1, v_2, \ldots, v_n\}$ and $V(G') = \{v'_1, v'_2, \ldots, v'_n\}$ such that the bijection $f: G \rightarrow G'$ defined as $f(v_i) = v'_i$ for all $i \in \{1, \ldots, n\}$ is an isomorphism, i.e. graph copies are identical for given orderings. Consider their disjoint union $L:= G \cup G'$ and connect each pair of vertices $v_i, v'_i$ by an edge with probability $q$. We call the resulting graph a graph generated by a {\em lateral random model} and refer to its automorphism $f$ as {\em left-right symmetry} (LR). Note that despite the similarity, this model is not an implementation of SBM for two communities.

\myparagraph{$k$-rewired LRM}
We further modify LRM model by rewiring random edges to different nodes to obtain more asymmetric graphs. The rewiring process simply takes a random edge, removes it, and adds a new edge to a random pair of non-connected nodes. If we consider $k$ consecutive rewiring, we refer to the resulting graph $k$-{\em rewired LRM}. Note that for constant values (or equivalently small values) of $k$ we can assume that the left-right symmetry $\LR$ of $k$-rewired LRM is the optimal approximate symmetry of this graph.

\begin{figure}[h!]
\begin{center}
    \includegraphics[width=0.5\textwidth]{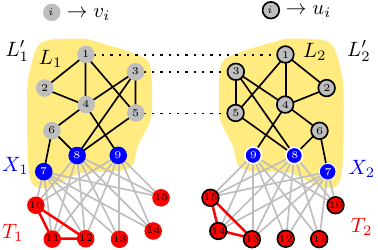}
	\caption{Example $(r,t)$-distorted LRM for $r = 3$ and $t = 6$. Original components $L_1, L_2$ have yellow backgrounds, sets $X_1, X_2$ are blue, and newly added twins (sets $T_1, T_2$) are red. The enumeration of nodes of each component is shown inside their shapes. Note that the enumerations of the two components are denoted using the the same progressions. We use node shape borders to distinguish them.}
	\label{fig:extendedLRM}
\end{center}
\end{figure}

\myparagraph{$t$-distorted LRM}
The initial evaluation of the algorithm shows that the performance is critically dependent on the initialization. One of the properties of the studied algorithms is their ability to achieve (a substantially improved) result when endowed with a suitable initialization, i.e. starting with a heuristic estimate of the targeted symmetry. To test this property, we develop another auxiliary test model that allows us to test the efficiency of finding an optimal solution when initialized with a semi-optimal one. First, for a given graph $G$, call two distinct nodes $u$ and $v$ {\em twins} if their neighbors in $G$ are the same, i.e. $N(u) = N(v)$. Note that when $u$ and $v$ are {\em twins}, a bijection $f$ that fixes all vertices of a graph $G$ except $u$ and $v$, for which $f(u) = v$ and $f(v) = u$, is an automorphism. To generate our test graph, let us first generate the LRM model $L$. Let us call the subgraphs of $L$ corresponding to the half-sized ER graphs $L_1$ and $L_2$. Recall that both components have mutually corresponding node orderings. Choose a small subset of nodes $X_1 \in L_1$ of size $r$ and select the corresponding (in terms of orderings) set $X_2 \in L_2$. Let $t \in \mathbb{N}$ be even. Generate two sets of nodes $T_1$ and $T_2$ each of size $t$. Extend the component $L_1$ into $L_1'$ as disjoint union $L_1' = L_1 \cup T_1$ and make every node in $T_1$ a twin in $L_1$ using set $X_1$ (add new edges so for each $x\in T_1$ holds that $N(x) = X_1$). In the same way, we extend the set $L_2$ to $L_2'$. Let the ordering of the vertices of $L_1'$ on the set $V(L_1)$ be the same as the original. Let the ordering of the remaining vertices corresponding to $T_1$ continue in an arbitrary way. Set the corresponding ordering of the $L_2'$ component in the same way. Add more edges to the graph so that the first $t/2$ vertices of $T_1$ form a complete graph. Next, add edges so that the second $t/2$ vertices of $T_2$ form a complete graph. We call the resulting graph $L'$ built on the above-described extended components $L_1'$ and $L_2'$ with $|X_1| = |X_2| = r$ and $|T_1| = |T_2| = t$ a $(r,t)$-{\em distorted LRM}. If the value of $r$ is known or unimportant from the context, such a graph can be called a $t$-{\em distorted LRM}. The whole construction is outlined in Figure~\ref{fig:extendedLRM}. 

Let us denote the ordering of vertices of $L_1'$ as $V(L_1') = \{v_1, v_2, \ldots, v_{n/2}\}$ and the ordering of vertices of $L_2'$ as $V(L_2') = \{u_1, u_2, \ldots, u_{n/2}\}$. Then we call an approximate symmetry swapping the vertices of $v_i$ for $u_i$ for all $i = \{1, 2, \ldots, n/2\}$ as $\LR$. Note that this permutation is not an automorphism but only an approximate automorphism with a defined number of errors, namely $2{t/2 \choose 2}$ pairs are different. Note that by correctly swapping the halves of the sets $T_1$ and $T_2$, we find the automorphism. 

\subsection{Data}\label{ss:data}

The human brain is a complex system that can be studied through diverse neuroimaging techniques, each focusing on distinct aspects of brain structure and function. We use the structural connectivity of the brain represented by the structure of white matter fibers. Such structure can be obtained by using diffusion-weighted imaging (DWI) and consecutive reconstruction (tractography) of white matter fibers connections~\cite{Zhang2021}. By partitioning the tractography outcomes into larger anatomical regions, structural interrelationships between corresponding anatomically relevant components (anatomical atlas) are inferred. This process results in a structural connectivity (SC) matrix that quantifies connection strengths between all pairs of regions.

For our study, we use the publicly available open dataset~\cite{Skoch2022} consisting of 88 healthy subjects' SC matrices with $90$ regions from Automated Anatomical Labeling (AAL) atlas~\cite{Tzourio-Mazoyer2002}. For more details on the dataset, see~\cite{Skoch2022}.
Finally, for the analysis, we binarize every SC matrix by considering only the $5\%$ of the strongest edges. Each graph generated by this procedure is connected.

\section{Results and analysis}
\label{sec:results}

\subsection{Simulations}

\myparagraph{General comparison of methods} 
This section describes a comparison of the achieved normalized approximate symmetry coefficients $S(A)$ obtained using either QSA or AFP method. We consider each run of the algorithm as a transformation of the initial permutation ($\IP$) into the final permutation ($\FinP$). Occasionally, we need to comment on the optimal solution. In this case, we use $\OPS$ to denote a set of optimal permutations. In general, we denote the value of the normalized approximate symmetry coefficient of the graph given by the matrix $A$ corresponding to a given permutation $P$ as $S(A, P)$ and, if the graph is known from context, as $S(P)$. Thus, for example, the value of the normalized approximate symmetry coefficient $S(A)$ when reaching the final permutation can be denoted as $S(\FinP)$.

We run the computations on Erd\H{o}s-Renyi (ER) graphs of three sizes ($20$, $50$ and $100$ nodes) and several edge densities from an interval $[0.1, 0.5]$. We conduct $100$ simulations for every combination of parameters (graph size and edge density) to estimate the distribution of $S(A)$. For every simulation, we generate the ER graph with the corresponding parameters. We run two methods for the comparison - AFP with the maximal number of fixed points set to half of the graph size and QSA.

\begin{figure}[h!]
\begin{center}
    \includegraphics[width=0.95\textwidth]{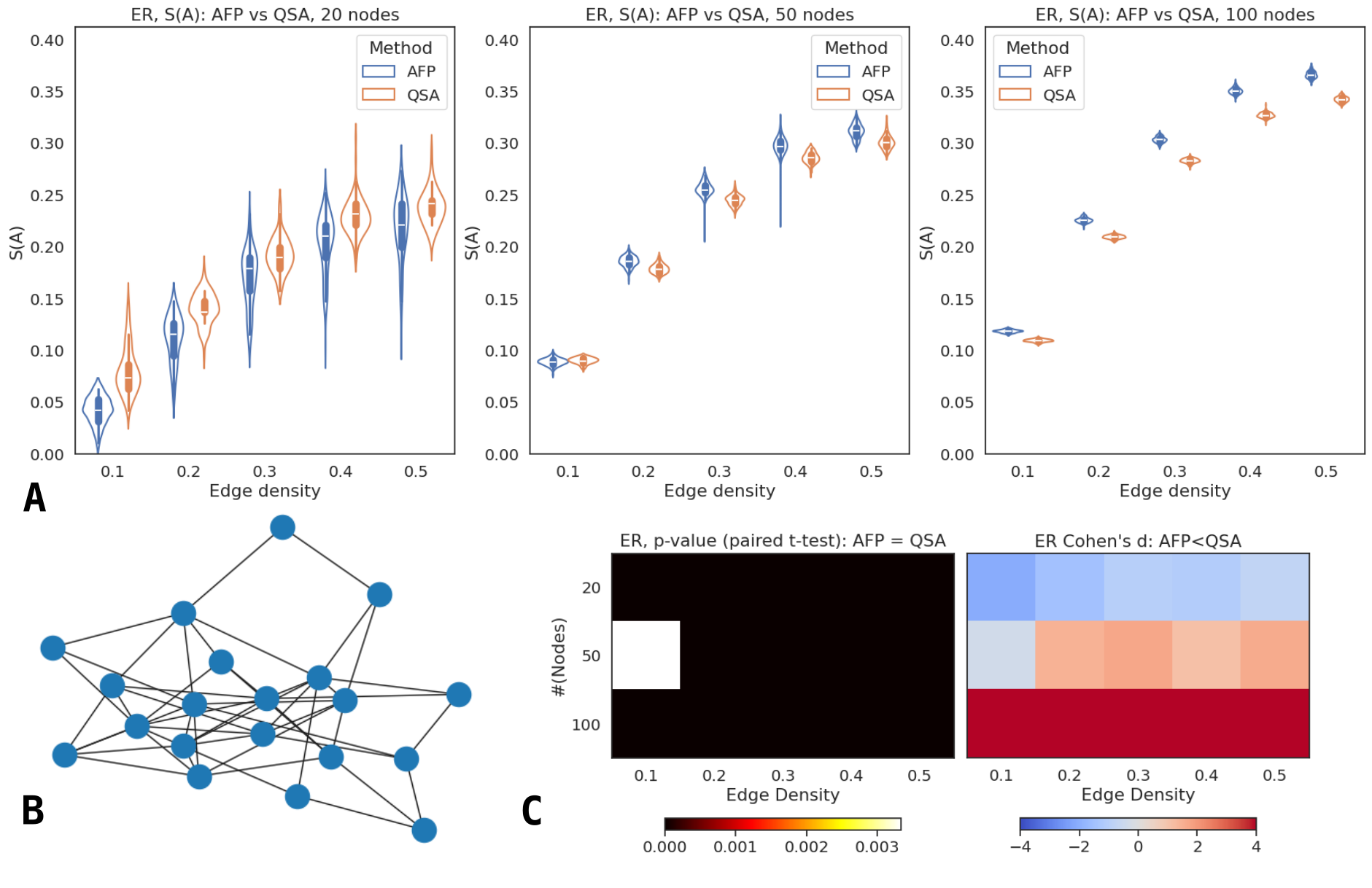}
	\caption{A. Comparison of $S(A)$ computed with Annealing with Fixed Points (AFP) and Quadratic Symmetry Approximator (QSA). Different graph sizes are shown on different subpanels ($20$, $50$ and $100$ nodes). Different edge densities are presented for every subgraph. B. Example of ER graph ($20$ nodes). C. Distributions comparison. For every graph size and edge density, we run the paired $p$-test to compare S(A) computed with AFP and QSA and compute Cohen's $d$ to estimate the effect size. }
	\label{fig:methodsComparison_ER_general}
\end{center}
\end{figure}

In Figure~\ref{fig:methodsComparison_ER_general}, the subpanel A shows the distributions of $S(A)$, obtained by AFP and QSA: the lower the values are, the closer the permuted graph is to the original one, and thus the corresponding permutation is closer to an automorphism. An example of a generated ER graph is presented in subfigure B. In subpanel C, we show the results of the statistical analysis. We ran a two-sided paired $t$-test (alpha was set to $0.05$) to compare the mean $S(A)$ estimated by AFP and QSA; corresponding $p$-values are shown in the left part of subpanel C. Black color corresponds to the graph characteristics, for which mean values of $S(A)$ distributions significantly differ for two methods (Bonferroni correction was applied to avoid the multiple comparison problem). As can be seen from the graph, $S(A)$ is different for most of the parameters. We also compute Cohen's $d$ (right graph on subpanel C), which indicates the direction of the difference and how large the effect size is. In our plot, the blue color corresponds to the situation when AFP is better than QSA (has lower $S(A)$ values), and vice versa for the red color. The more intense the color, the larger the difference we observe between the $S(A)$ mean values. 

It is clear from the results that the behavior of the methods heavily depends on the graph parameters. From our simulations, AFP performs better for the small ($ 20 $ nodes in our case) or average-sized but sparse (up to $ 0.1 $ edge density) graphs. However, for the average and larger graph sizes, QSA outperforms the AFP in terms of $S(A)$. We obtained similar results for other graph models (grid, BA, SBM); the results are presented in the Appendix in Figures~\ref{fig:addSim_grid}, \ref{fig:addSim_BA}, \ref{fig:addSim_SBM}. 
We conjecture that for smaller graphs, the slight dominance of AFP is due to the relatively low combinatorial complexity of potential automorphisms (the simplicity of AFP then becomes an advantage). As the graph grows, QSA takes advantage of the more efficient search.

Since for the smaller graph sizes considered, the comparison of methods is burdened with relatively less combinatorial complexity of the problem, to test the performance in a more realistic yet demanding setting, we will consider only graphs of size $200$ in the following to test the algorithms in a demanding, yet realistic setting.

\begin{figure}[h!]
\begin{center}
    \includegraphics[width=0.95\textwidth]{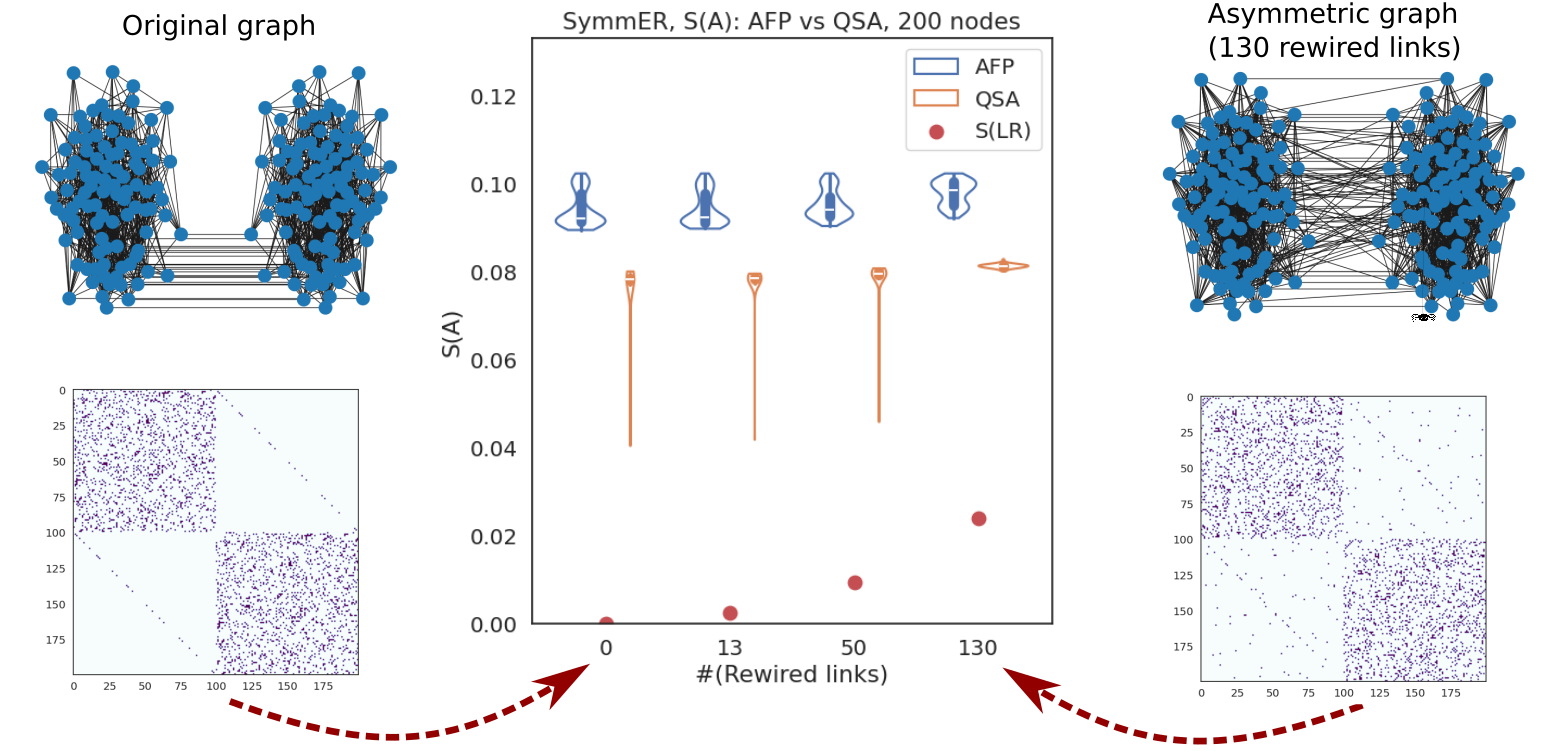}      
	\caption{Comparison of S(A) distributions obtained by AFP and QSA on graphs generated by LRM (with perfect symmetry) and rewired LRM (with approximate symmetries given by increasing numbers of rewired links). Original (LRM) and the most asymmetric graphs ($k$-rewired LRM with maximal $k=130$) together with corresponding adjacency matrices are shown in the figure. }
	\label{fig:methodsComparison_symER}
\end{center}
\end{figure}

\myparagraph{Symmetric graphs and initialization}
In this section, we use a random lateral model (LRM) $L_{200, 0.15,0.25}$ having one dominant symmetry -- referring to section~\ref{s:models}, called left-right symmetry ($\LR$). To test graphs having only approximate symmetries, we further generated $k$-rewired LRMs. This way, we obtained a sequence of graphs from completely symmetric to graph with $k = 130$ rewires from the symmetry. Then, we used both methods to obtain the approximate symmetries measured by $S(A)$. Algorithms were run $75$ times each. We compared $S(A)$ values in Figure~\ref{fig:methodsComparison_symER}.

The results shown in Figure~\ref{fig:methodsComparison_symER} show that AFP does not find the optimal symmetry for any of the versions of graphs; its average $S(A)$ value for the symmetric graph is approximately $0.09$, which corresponds roughly to $900$ mistakes (as evaluated by $E(A)$). For QSA, the result is better (the best reached $S(A)$ value is $0.04$, approximately $400$ mistakes), but for most of the simulations, resulting approximate symmetries are far from the optimal ($\LR$) permutation (QSA found permutations with $S(A)$ around $0.08$, approximately $800$ edges off). Therefore, as the next exploration step, we run the algorithms with a meaningful initialization instead of a random permutation.

Except for running the methods with an optimal $\LR$ initialization, we also run the algorithms with initial permutations that slightly differ from $\LR$. We construct such initialization from $\LR$ using the following procedure (defined for general starting permutation $P$). Start with a permutation $P$. Next, uniformly select two nodes and swap their images under permutation. We repeat this procedure $\ell$-times and call the resulting permutation $\ell$-{\em reshuffled $P$} and call each step a {\em swap}. Specifically, we will work with $\ell$-reshuffled LR. This allows us to explore the ability of the algorithm to find an optimal solution while the algorithm is initialized with a permutation not far from the optimal one. 

\begin{figure}[h!]
\begin{center}
    \includegraphics[width=0.95\textwidth]{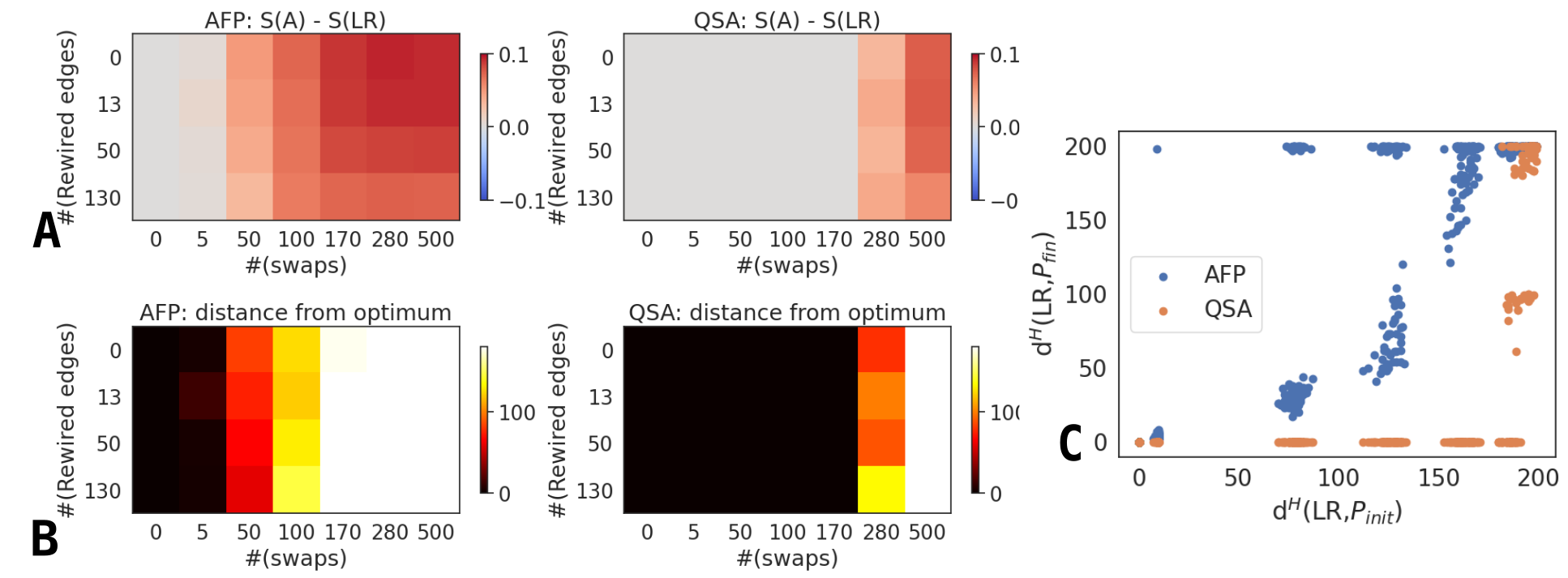}     
	\caption{Comparison of the performance of the algorithms when running on $k$-rewired LRM and initialized with (semi-)optimal permutation ($\ell$-reshuffled $\LR$ for $\ell=\{0,\ldots, 500\}$ swaps). 
    Panel A shows the difference between an algorithm's solution $S(A)$ and an optimal solution $S(\LR)$ ($\LR$ symmetry). Panel B shows the Hamming distance between an optimal ($\LR$) and permutation provided by the respective algorithm ($P$), i.e. $\HD(\LR, P)$. Panel C shows the relationship between the initial permutation $\IP$ and the final permutation $\FinP$ in terms of Hamming distance. More specifically, the scatter plot between Hamming distances  $\HD(\LR, \IP)$ and $\HD(\LR,\FinP)$; $\HD$ stands for Hamming Distance. For every set of parameters, $9$ independent simulations were computed; panels A and B show the averaged result.}
	\label{fig:methodsComparison_symER_LRinit}
\end{center}
\end{figure}

Figure~\ref{fig:methodsComparison_symER_LRinit} compares AFP with QSA for $k$-rewired LRM using $\ell$-reshuffled $\LR$ initialization ($\LR$ with different amounts of swaps). Panels A and B have similar structures: for both methods, on the $x$-axis, we show the number of swaps ($\ell$) of the $\ell$-reshuffled $\LR$ initialization (error in the initial symmetry estimate), and on the $y$-axis, the number $k$ of rewired edges for $k$-rewired LRM (the deviation of the networks from perfect symmetry). 
Panel A shows a difference between an approximate symmetry $S(A)$ obtained after running the algorithm with the semi-optimal initialization and an approximate symmetry $S(\LR)$ of $\LR$ permutation, which we assume to be the optimal permutation, i.e. $\LR \in \OPS$. Since the goal is to have an approximate symmetry value as low as possible, the best result would be zero difference. A positive difference means that the algorithm did not find a solution as good as the optimal one. Panel B shows the Hamming distance between the $\LR$ and the final permutation $\FinP$, i.e.  $\HD(\LR, \FinP)$.

The performance of methods depends on how close the initial solution is to the optimal one; for the slight deviation (small amount of swaps), both methods find the optimal solution. For the more significant deviation, methods show larger values of $S(A)$ compared to the optimal solution. The number of rewired edges $k$ does not have a strong influence, if any. 
The difference between $S(A)$ and $S(LR)$ is smaller for relatively larger numbers of rewires $k$. However, that happens because $S(LR)$ grows with increasing $k$ for $k$-rewired LRM, while $S(A)$ stays almost constant.
On the other hand, the number of swaps $\ell$ for the initialization permutation seems to have a larger effect. This is most likely due to the fact that we affect the search for the minimum itself by providing less relevant initial localization. This may eventually divert the search from the correct direction. In both cases, however, QSA performs better than AFP. More specifically, QSA has a larger range of parameters (mainly in the number of $\ell$-swaps) for which it successfully finds a path to the optimal $k$-approximate symmetry.
Figure~\ref{fig:methodsComparison_symER_LRinit} panel C shows the scatter plot between the Hamming distances $\HD(\IP,\LR)$ and $\HD(\FinP,\LR)$. Ideally, we expect all the points to lay on the horizontal line corresponding to $0$ distance from the $\LR \in \OPS$. However, we see the increase for both methods, for AFP, however, much earlier than for QSA.

\begin{figure}[h!]
\begin{center}
    \includegraphics[width=0.95\textwidth]{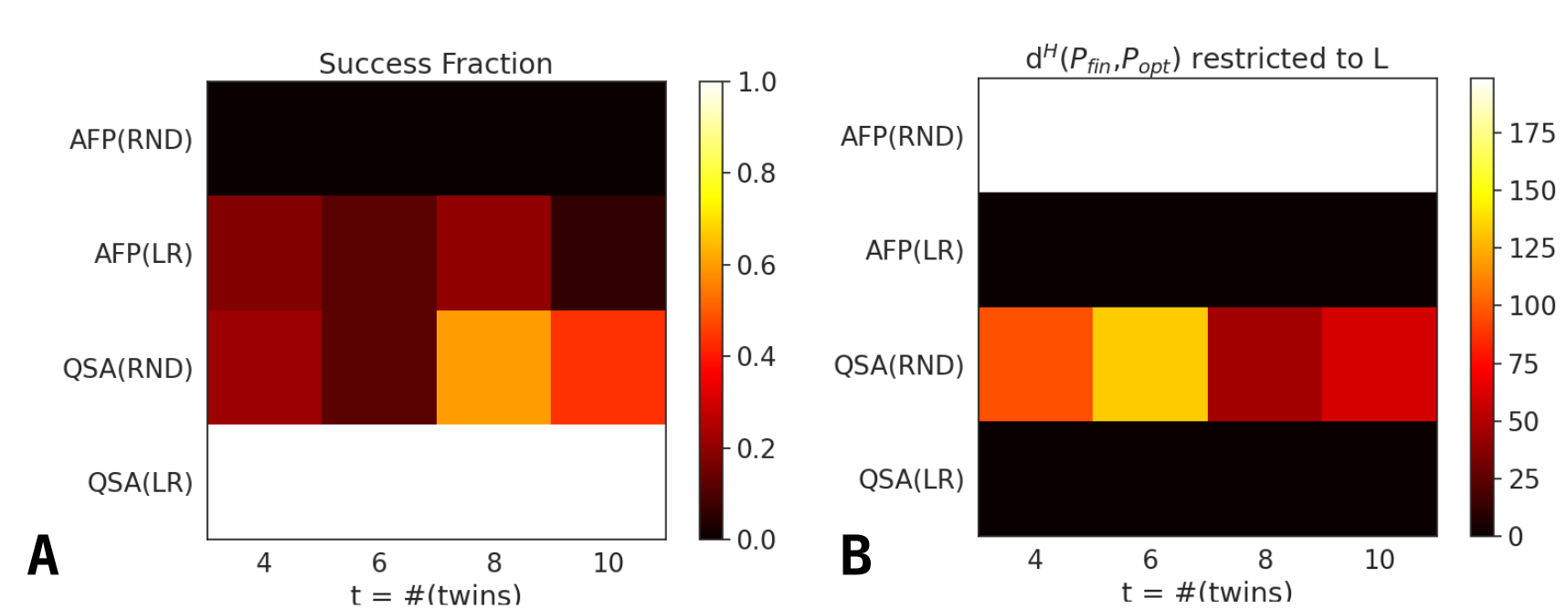}
	\caption{
    Panel A shows the fraction of simulations converging to the perfect symmetry for different amounts of twins. Panel B shows the Hamming distance between the final and optimal permutations, restricted to $V(L)$. $P_{\text{opt}}$ is one of the optimal permutations, i.e. $P_{\text{opt}} \in \mathcal{P}_{\text{opt}}$}
	\label{fig:methodsComparison_symER_leaves}
\end{center}
\end{figure}

In the next part, we use the graphs generated from the t-distorted LRM. The goal is to make use of a situation where we know exactly how far we are from the optimal (in this case, real) symmetry. 
The advantage of the t-distorted LRM model is that it gives us during the testing full control over how far the initial LR permutation is from the optimal symmetry, and we can thus assess the effect of the error of the initialization (distance of initialization from the optimal solution).
Note that in the case of the t-distorted LRM, the optimal symmetry roughly corresponds to the $\LR$ permutation given by the orderings on the respective components of $L$, namely $L_1$ and $L_2$, with a small modification on the set $T = T_1 \cup T_2$. Hence, one can expect that when initialized by the LR permutation, the algorithms would gradually modify the initial permutation on a set $T$ while keeping operations of $\LR$ permutation on the subgraph $L$ unchanged. Since the nature of the optimum of the $t$-distorted LRM can be precisely modified by the parameter $t$, we can test the ability of the algorithms to find the optimum in a gradually modified and localized form. 

The graphs generated here by the $t$-distorted model are always of size $200$ plus $2t$ twins, and we run $50$ simulations for each setting. The ratio of successful algorithm runs across simulations, giving the correct symmetry, is shown in Figure~\ref{fig:methodsComparison_symER_leaves}, subgraph A. These results show the clear dominance of the QSA method. Moreover, thanks to the utilization of the $t$-distorted LRM, we can also observe how efficient the algorithms are in achieving the target optimal permutation, which is $\LR$, except for the subgraphs given by the twins. These results are shown in Figure~\ref{fig:methodsComparison_symER_leaves}, subgraph B, showing that if we appropriately initialize both AFP and QSA, they preserve the LR permutation on subgraph $L$. However, QSA succeeds better in reconstructing the correct permutation of vertices on the $T_1 \cup T_2$. For random initialization, the AFP method completely fails to find the optimal solution. The resulting solutions provided by this method are not even able to find the $\LR$ permutation on $L$. When considering the random initialization for the QSA method, part of the simulations resulted in the optimal permutation.

\subsection{Brain}

\begin{figure}[h!]
\begin{center}
    \includegraphics[width=0.75\textwidth]{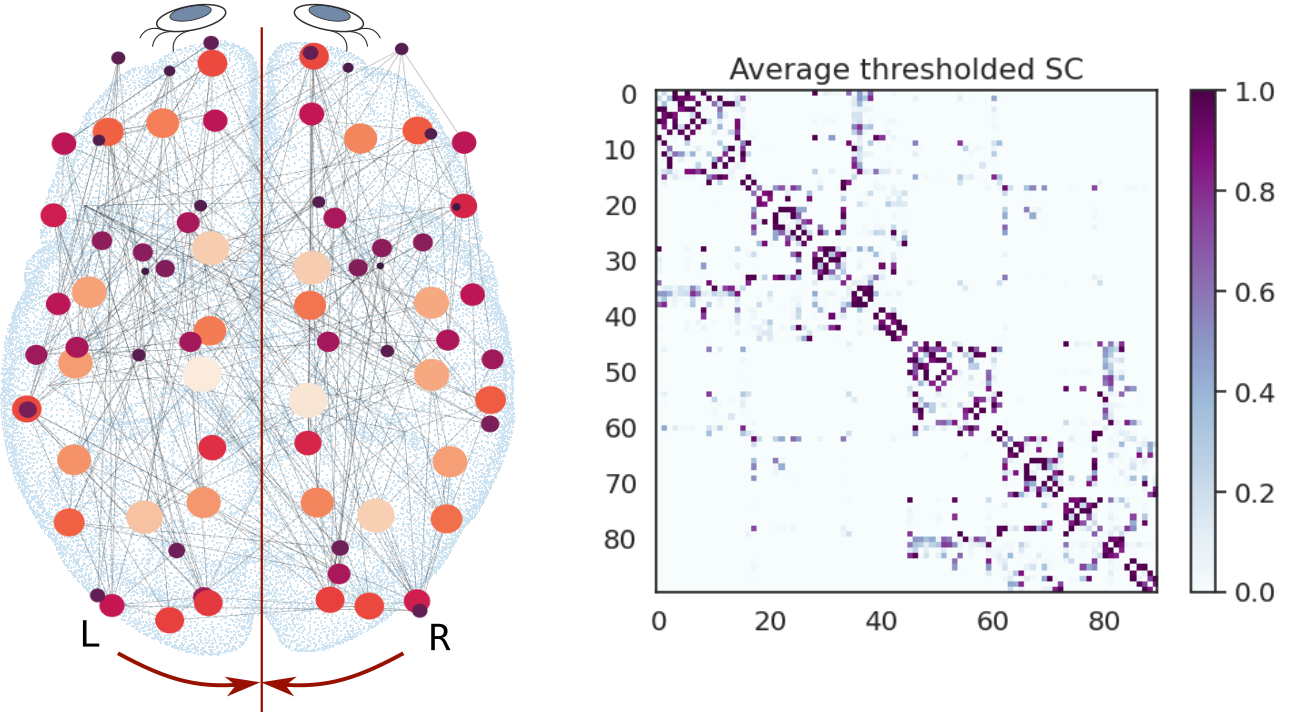}
	\caption{Left: The approximate position of brain regions we use as graph nodes, their connectivity in a representative subject, and the corresponding average SC. The size and color of a node correspond to the z-axis that is not visible on a graph: the larger and lighter the node is - the higher it is in reality. Right: Average SC matrix, thresholded to the density of 5 percent.}
	\label{fig:brain_basic}
\end{center}
\end{figure}

Apart from the simulated data, we also analyzed the human brain structural connectivity (SC). For every subject, we work with the binarized SC matrix. As explained in the Data subsection, its $90$ nodes correspond to brain regions from the AAL atlas. Their approximate positions are shown in Figure~\ref{fig:brain_basic}; here, $x$ and $y$ coordinates approximately correspond to the centers of mass of the respective brain regions in the horizontal plane of the template anatomical atlas space, while the color and size of the vertex correspond to $z$ (vertical) axis, which can not be shown on a $2D$-picture. Connections between the nodes were taken from an example subject.
Figure~\ref{fig:brain_basic}B shows an average structural connectivity for all $88$ subjects. 
The first $45$ regions correspond to the left hemisphere, the second $45$ regions to the right hemisphere.

Based on the neurophysiology, we assume that the brain is approximately symmetric, at least in the sense of lateral symmetry~\cite{Gunturkun2020}; however, unlike in the case of a symmetric LRM, we can not assume that it is perfectly laterally symmetric (the $\LR$ permutation is not optimal)~\cite{Jajcay2022}. Therefore, we perform our computations to explore brain symmetry. 

\begin{figure}[h!]
\begin{center}
    \includegraphics[width=0.75\textwidth]{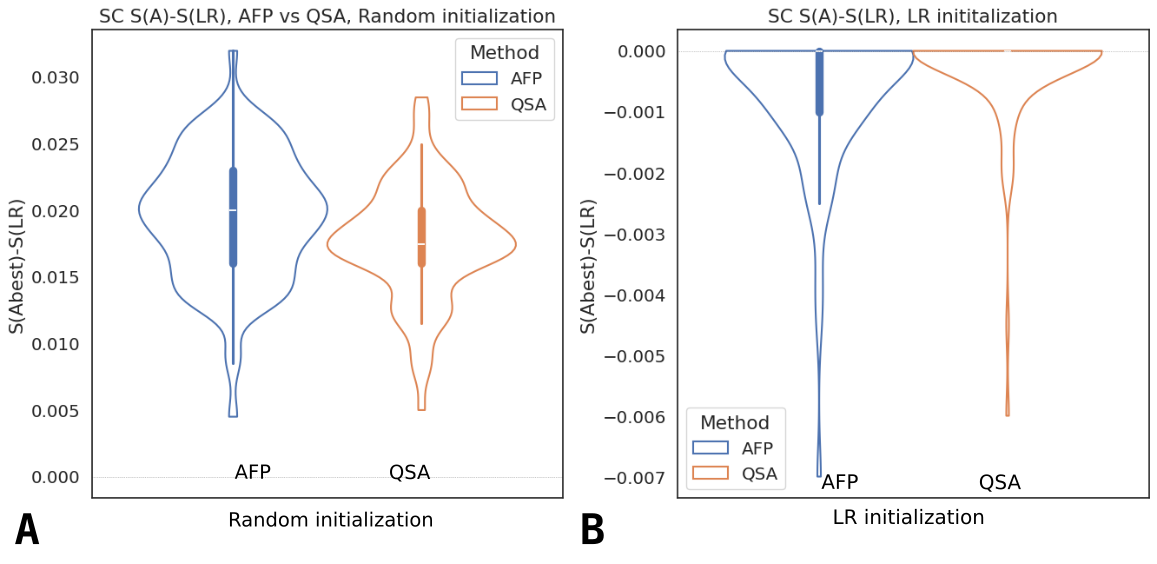}
	\caption{Panel A shows distributions of differences $S(A)-S(LR)$, obtained by AFP and QSA on the brain with the random initialization, best simulation ($A_{\text{best}}$) of 9 runs. Panel B shows distributions of differences $S(A)-S(LR)$, obtained by AFP and QSA on the brain with the LR initialization, best simulation ($A_{\text{best}}$) of 9 runs.}
	\label{fig:brain_init}
\end{center}
\end{figure}

First, we run both algorithms with a random initialization and compare the resulting permutation with the $\LR$ permutation. The result can be found in Figure~\ref{fig:brain_init}, panel A. 
We run both algorithms $30$ times per subject and select the best run for each subject; the distribution across subjects is shown in the figure. Even though QSA results are significantly better than those of AFP, both methods provide results that are worse that those provided by the LR permutation. Not being able to outperform the heuristic LR permutation suggests that finding the true approximate symmetry for the brain would be even more challenging for both methods. Thus, we attempt to improve the results by initializing the methods with expert knowledge of the left-right approximate symmetry.

The results of the runs after initializing LR permutations are shown in Figure Figure~\ref{fig:brain_init}, panel B. From these results, it seems that if we initialize the approximate symmetry in the brain network using the $\LR$ permutation, AFP is slightly better than QSA in finding a solution that is more symmetric than $\LR$. This is a different result than was shown in the LRM model. Thus, it seems that the brain model has a specific structure that may be aiding the otherwise less efficient AFP in improving approximate symmetry estimation. 
This is likely due to the specific structure of both the final permutation and the left-right approximate symmetry itself.

\begin{figure}[h!]
\begin{center}
    \includegraphics[width=0.95\textwidth]{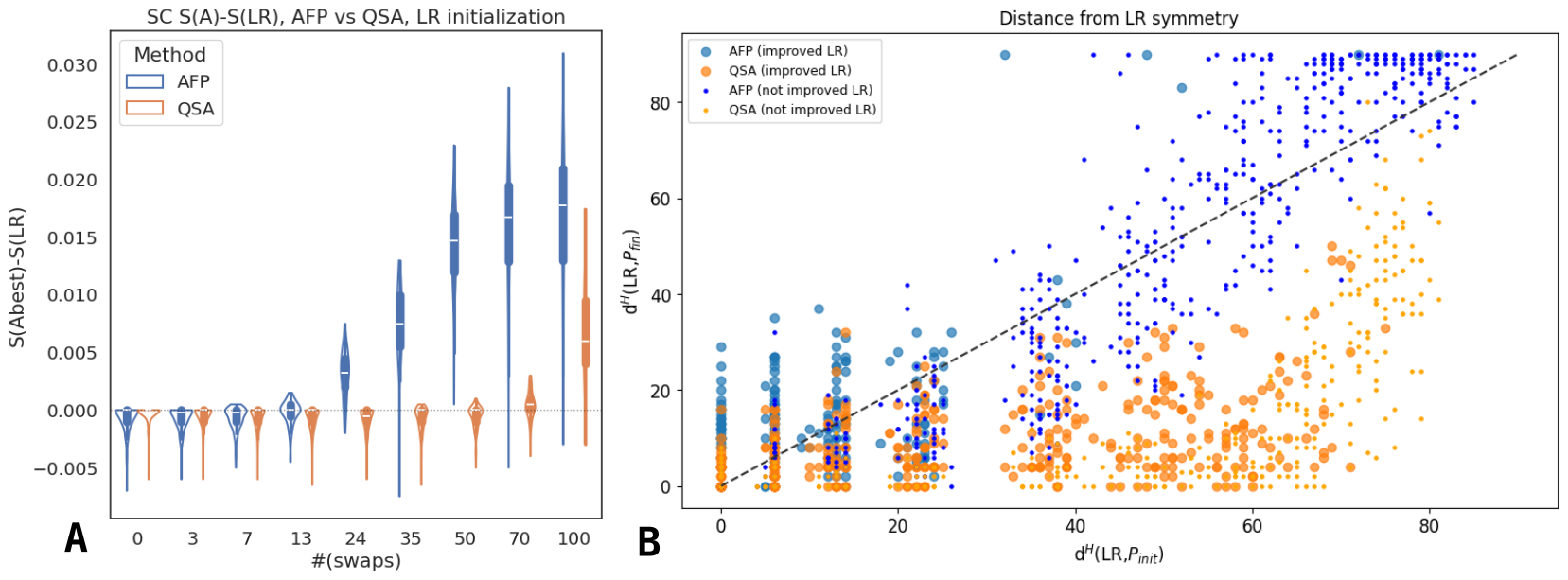}
	\caption{Panel A shows results of runs for $\ell$-rewired LR initialization for different $\ell$ (swaps). Panel B shows the Hamming distances $\HD(\LR, \IP)$ and $\HD(\LR, \FinP)$ for both algorithms and for those runs that improved LR permutation (larger points) that those that do not (small points).}
	\label{fig:brain_init_rewired}
\end{center}
\end{figure}

Thus, we also test the case when the initialization is not directly $\LR$ but is a $\ell$-reshuffled version of it. This will test situations where we do not give the algorithm exactly a particular natural symmetry but an inaccurate estimate. The question is whether AFP retains its advantage in the case of gradual drift away from the $\LR$ initialization. As previously, we repeat the procedure $30$ times and select the best result for every subject. 
We present the resulting distributions in Figure~\ref{fig:brain_init_rewired}, panel A. We show the difference between the obtained symmetry and the one corresponding to the LR symmetry, i.e. $S(A) - S(\LR)$.

We observe that when starting strictly at $\LR$, then for most subjects, both algorithms improve the $S(LR)$ value. When moving away from $\LR$ symmetry (swaps), AFP starts failing to find a permutation that would improve $\LR$, and QSA performs well up to a significant distance between an initial permutation and $\LR$.
Panel B of Figure~\ref{fig:brain_init_rewired} shows how far initial and final permutations are from LR for those runs when the original LR permutation was (or was not) improved in terms of $S(A)$. On top of what we already know from panel A, we also see that QSA tends to go back closer to $\LR$ symmetry even when started far enough. This statement can be supported by finer filtering of the results from Panel B. From the results of this plot, we select only those where the approximate symmetry value has been improved (marked as larger dots in Panel B of Figure~\ref{fig:brain_init_rewired}). 

\begin{figure}[h!]
\begin{center}
    \includegraphics[width=0.95\textwidth]{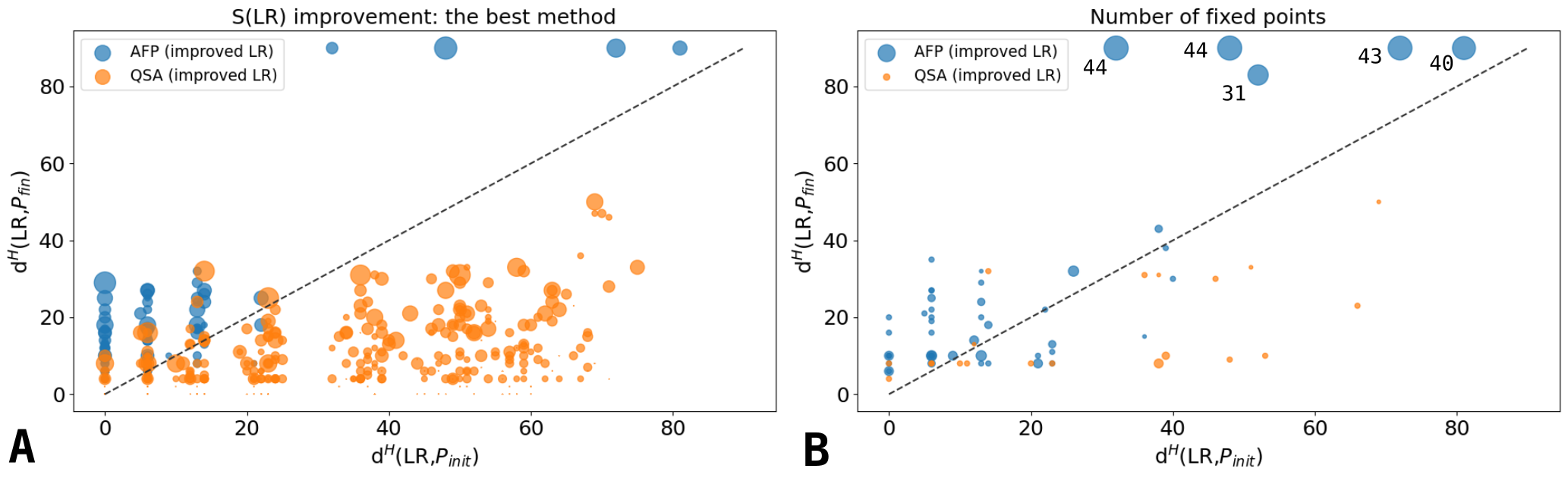}   
	\caption{Panel A shows the scatter plot between Hamming distances $\HD(\LR, \IP)$ and $\HD(\LR, \FinP)$ for both algorithms and those runs that improved LR permutation. The size of the data points scales with the improvement (the larger the data point, the greater the improvement). Panel B redraws the data points from Panel A, but this time, the size of the points corresponds to the number of fixed points of the solution. Numbers close to the points correspond to the number of fixed points.}
	\label{fig:brain_distrib}
\end{center}
\end{figure}

Furthermore, for each initialization in which there was at least one improvement for  the AFP or QSA method, we selected the result of the method that performed better. The result of this filtering can be seen in Figure~\ref{fig:brain_distrib}, panel A. It can be seen from the figure that QSA performs much better when the initial permutation is not directly LR. Moreover, this method tends to return solutions rather close to the natural LR symmetry. On the other hand, AFP, which succeeds mainly for initializations closer to LR, tends to generally search for solutions more distant from this permutation.

Interesting results are also the solutions to the problem found using the AFP method, which exhibit a very large distance to LR symmetry. These results, shown in the upper part of Panel A in~Figure~\ref{fig:brain_distrib}, show that approximate symmetries better than LR and very far from LR can be found in some subject structural connectivities. Upon expectation of the number of fixed points of the given results (data not shown), we observed that these results represent permutations that contain a large number of fixed points. This indicates their very specific structure. It can also be seen from these results that the QSA method tends to avoid fixed points, whereas the AFP method uses them. This is one consequence of our modification of the original annealing approach via the inclusion of fixed points. In addition to the nature of the methods, this also suggests that the brain contains different types of approximate symmetries. Classification and description of these symmetries is a possible area for future work.

\section{Discussion and conclusions}
We revisit an approximate network symmetry problem, first raised in~\cite{liu2020approximate}. We improve the previously proposed method based on simulated annealing by introducing the possibility of considering fixed points, giving rise to the Annealing with Fixed Points (AFP) approach. We also proposed a new alternative method (QSA) based on a modification of the approach for testing approximate isomorphism~\cite{vogelstein2015fast} (QAP). We numerically tested the proposed methods on different types of random networks representing different aspects of real complex networks, including the Erd\H{o}s-Renyi, Barab\'{a}si-Albert, or Stochastic block model. In all models, the QSA algorithm showed better performance than AFP in the random initialization case. The challenge remains concerning achieving a true optimum, which is difficult to assess in the tested networks (the ground truth is not known in these cases). 

Some real-world systems have intrinsic symmetry given by the nature of the system. An example is the lateral symmetry of the brain. However, such symmetry may not be an exact symmetry in the system but only an approximate one. It may not even be the best approximate symmetry, both due to the asymmetries of the real system and observation errors, thus warranting the search for an alternative optimum. The typical situation in these cases is that the optimal symmetry is not too far from the symmetry given by the nature of the system. A technical problem was how to test algorithms when they are instantiated by a permutation near the target. For this purpose, we defined several graph models reflecting the properties of systems described above (with an a priori known approximate symmetry).

The first model is a lateral random model (LRM) inspired by the symmetry of brain networks. This model produces a single dominant symmetry ($\LR$ symmetry) in the random graph. Its variant, called $k$-rewired LRM, allows modifying the LR symmetry to be an approximate symmetry but still most likely the best of all approximate symmetries. We can use this model to test runs of algorithms with different initializations around $\LR$ and test its ability to reach the optimum. Note that this optimum is either exact or approximate (with rewired edges). Testing variants of this model show a significant dominance of the QSA method. 

The rewired version of the LRM model has a clearly defined approximate symmetry using the $\LR$ permutation. However, deviations from this symmetry are random and poorly localized. To test the algorithm's ability to localize changes in the initial permutation correctly, we created another model called $t$-distorted LRM. In this model, we expect to preserve part of the initial permutation and modify a selected segment. While testing this model, QSA was able to find the optimum in all instances. In contrast, AFP, although it preserved the part of the permutation intended to be fixed, was not able to correctly determine the altered part. In the situation where random initialization was used, the AFP approach was completely unsuccessful. On the other hand, QSA was able to find an optimal solution for a substantial proportion of the tested input permutations.

Evaluation of approximate symmetries on the brain network given by structural connectivity yielded results indicating non-triviality of the structure of the corresponding network. First, the $\LR$ permutation does not appear to be an optimal symmetry of the brain network, as shown consistently by both methods. Although the resulting permutations were not fundamentally far from $\LR$, improvements were found for most subjects. A surprising result is that, in contrast to the clear result for the LRM variants, AFP seems to be slightly better if exactly LR permutation was used for initialization. This suggests an interesting network structure that cannot be modeled by the aforementioned models. When we tried to run the algorithm for initialization more distant from $\LR$, the superiority of the QSA method again became apparent, and moreover, the algorithm found approximate symmetry better than the LR permutation, even for quite distant initial permutations. Moreover, from a certain threshold modification of the initial LR permutation, AFP found essentially no improvements.

Another interesting result is that the arbitrary improvement of the LR permutation was essentially always below a defined distance limit from the LR. Surprisingly, there were a few exceptions found by the AFP method. These results had a large distance from the original LR symmetry, showing a second extreme with some subjects having very distant approximate symmetries better than LR. It is also worth noting that these solutions had a large number of fixed points and finding them using AFP was made possible by extending it to include these fixed points. These results suggest an interesting symmetry structure around the LR permutation that will certainly be the subject of further investigation.

\section{Declarations}

\subsection{Conflicts of interest}
The authors have no conflicts of interest to declare that are relevant to the content of this article.

\subsection{Data availability statement}
As mentioned in Section~\ref{ss:data} the studied dataset is publicly available and open~\cite{Skoch2022}. 

\subsection{Ethics approval and consent to participate}
The publicly available and open dataset used in this study comes from study~\cite{Skoch2022} that was conducted in accordance with the Declaration of Helsinki. The local Ethics Committee of the Prague Psychiatric Center approved the protocol on 29 June 2011 (protocol code 69/11). All participants were informed about the
purpose of the study, the experimental procedures, as well as the fact that they could withdraw from the study at any time, and provided written informed consent prior to their participation.

\section*{Acknowledgements}
The authors were supported by the Czech Science Foundation Grant No. 23-07074S.

\bibliographystyle{abbrv}
\bibliography{biblio2.bib}

\section*{Appendix A Other simulations}

We compare AFP and QSA on several different random models and sets of parameters. We set the maximum number of the fixed points for AFP to half of the graph size. For every model, we test graphs of size 50, 100 and 150 nodes. Then, for every graph size, we explore several model-specific parameters. For every setup, we compare the distributions of $S(A)$ obtained by AFP and QSA. The structure of the results is the same as in Figure~\ref{fig:methodsComparison_ER_general}: panel A demonstrates $S(A)$ distributions for AFP and QSA for all three graph sizes. Panel B shows an example of a graph of a particular model. Panels C and D represent the statistical analysis: $p$-value for paired two-sided $t$-test and Cohen's $d$ for all the parameter settings. Color scales are similar to Figure~\ref{fig:methodsComparison_ER_general}. For the $p$-values, 
the dark color corresponds to the significant difference between the distributions after the Bonferroni correction. For Cohen's $d$, the blue color corresponds to the setting where AFP is better, and the red color indicates the QSA win.

\begin{figure}[h!]
\begin{center}
    \includegraphics[width=0.75\textwidth]{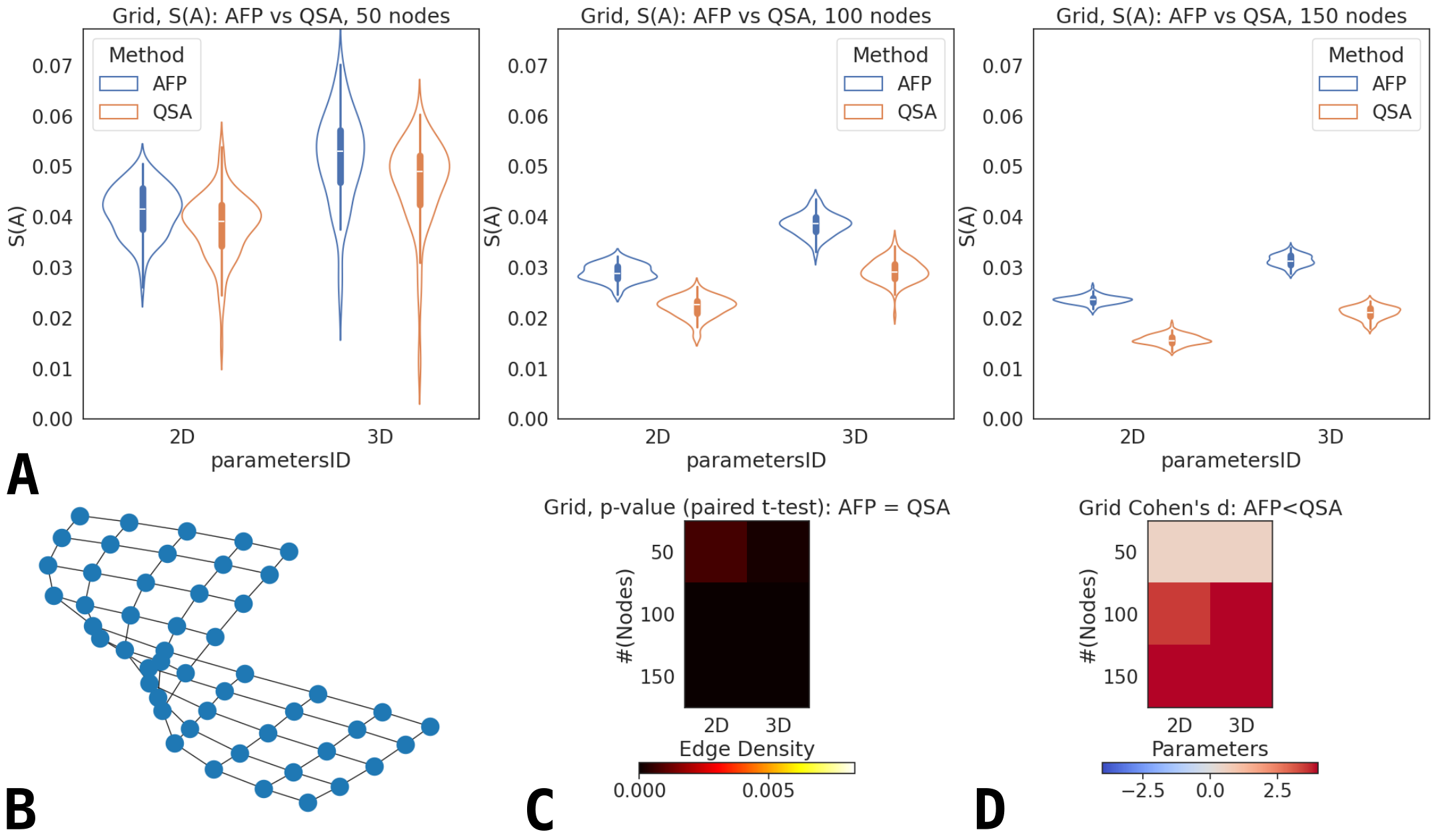}
	\caption{Grid. Panel A demonstrates $S(A)$ distributions for AFP and QSA for all three graph sizes. Panel B shows an example of a grid. Panels C and D represent the statistical analysis: $p$-value for paired two-sided $t$-test and Cohen's $d$ for all the parameter settings. }
	\label{fig:addSim_grid}
\end{center}
\end{figure}

Figure~\ref{fig:addSim_grid} shows the analysis on a grid. As a model-special parameter, we use the grid dimensionality (2D and 3D). For the 2D grid, its size is set to $5\times x$, where $x$ is varied to get $50$, $100$ and $150$ nodes. Similarly, for the 3D grid, it is always $5 \times 2 \times x$, where $x$ is varied to get the desired graph size. 

\begin{figure}[h!]
\begin{center}
    \includegraphics[width=0.75\textwidth]{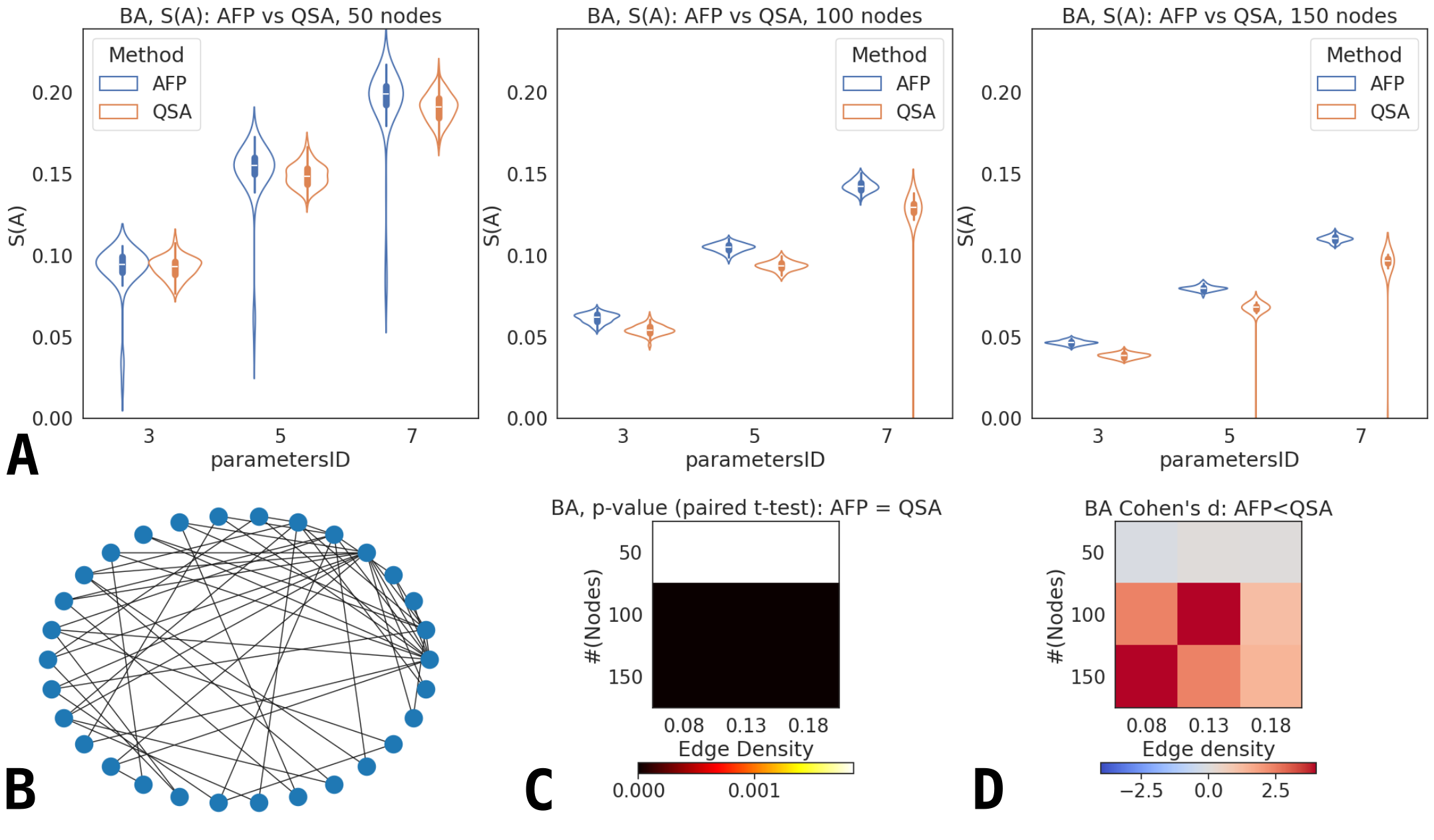}
	\caption{Barabási-Albert (BA) model. Panel A demonstrates $S(A)$ distributions for AFP and QSA for all three graph sizes. Panel B shows an example of a grid. Panels C and D represent the statistical analysis: $p$-value for paired two-sided $t$-test and Cohen's $d$ for all the parameter settings.}
	\label{fig:addSim_BA}
\end{center}
\end{figure}

Figure~\ref{fig:addSim_BA} shows the Barabási-Albert model. On panel A, as a model-specific parameter, we select the number of edges to attach from a new node to existing nodes. For panels C and D, the edge density is used on the $x$-axis. Note that a lower number of edges to attach to the new node corresponds to the lower edge density. 

\begin{figure}[h!]
\begin{center}
    \includegraphics[width=0.75\textwidth]{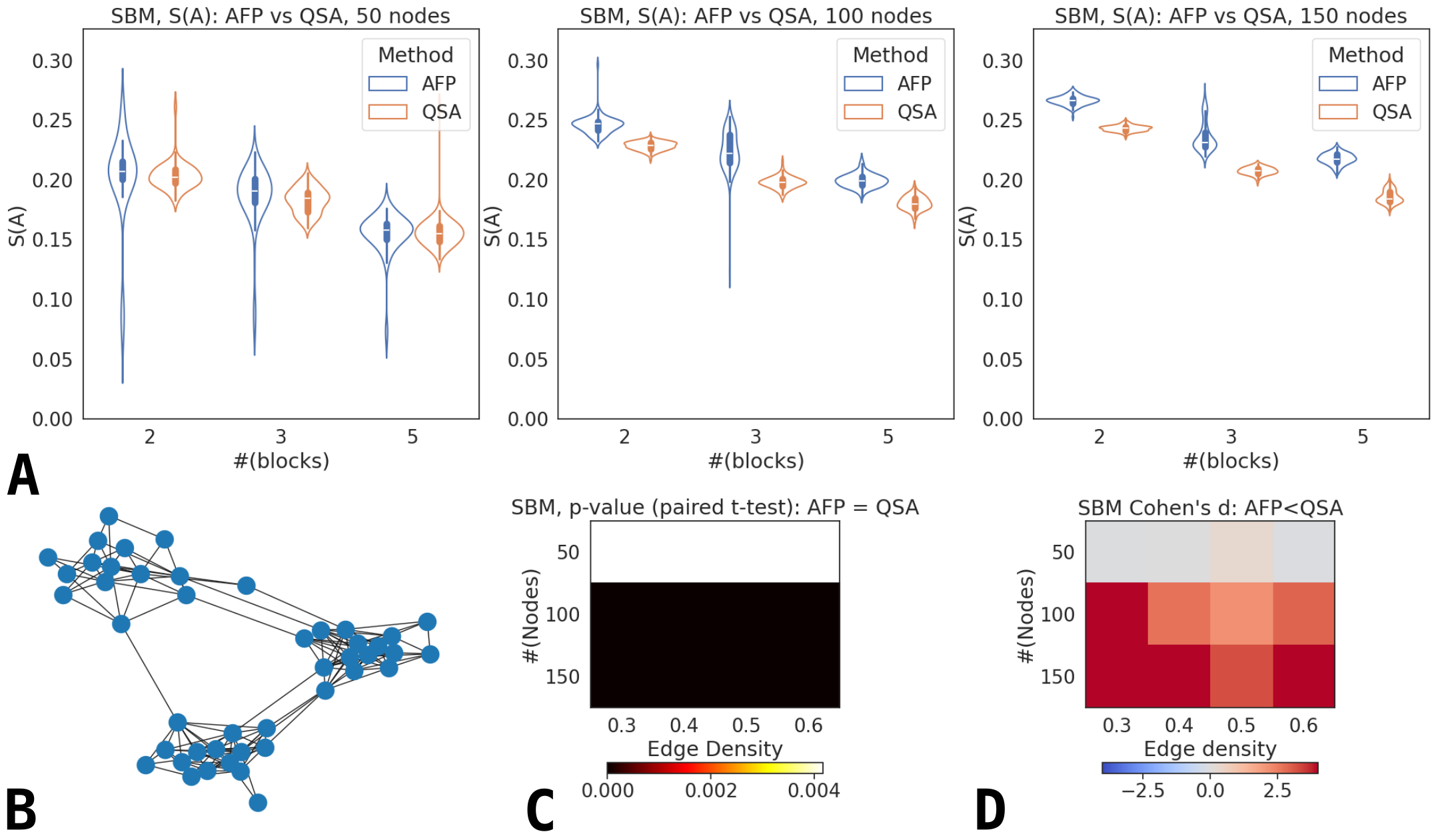}
	\caption{Stochastic Block Model (SBM). Panel A demonstrates $S(A)$ distributions for AFP and QSA for all three graph sizes. Panel B shows an example of a grid. Panels C and D represent the statistical analysis: $p$-value for paired two-sided $t$-test and Cohen's $d$ for all the parameter settings.}
	\label{fig:addSim_SBM}
\end{center}
\end{figure}

Figure~\ref{fig:addSim_SBM} shows the Stochastic Block Model with 2, 3 and 5 blocks. Probabilities inside blocks are set to 0.5, and the probabilities between blocks are set to 0.1. As a consequence, the number of blocks is anticorrelated with the edge density. 

For all the models and all the graph sizes, QSA is comparable or performs better than AFP; for larger graphs, the tendency is more pronounced. 

\end{document}